\documentclass[12pt]{article}
\usepackage{jheppub}

\usepackage{color}

\definecolor{myZero}{rgb}{0.858, 0.188, 0.478}
\definecolor{LinearBox}{RGB}{51, 102, 0}
\definecolor{LtGr}{RGB}{100,252,100}
\usepackage{exscale,latexsym}
\usepackage{amsmath}
\usepackage{cite}

\def\Year{\expandafter\eatPrefix\the\year}
\newcount\hours \newcount\minutes
\def\monthname{\ifcase\month\or
January\or February\or March\or April\or May\or June\or July\or
August\or September\or October\or November\or December\fi}\def\shortmonthname{\ifcase\month\orx
Jan\or Feb\or Mar\or Apr\or May\or Jun\or Jul\or
Aug\or Sep\or Oct\or Nov\or Dec\fi}

\def\TimeStamp{\hours\the\time\divide\hours by60%
\minutes -\the\time\divide\minutes by60\multiply\minutes by60%
\advance\minutes by\the\time%
${\rm \shortmonthname}\cdot   \if\day<10{}0\fi\the\day\cdot   \the\year
\qquad\the\hours:\if\minutes<10{}0\fi\the\minutes$}

\newskip\humongous \humongous=0pt plus 1000pt minus 100pt

\newif\ifdtup

\newcounter{eqnumber}[section]

\def\PT{C_{PT}}
\def\CY{Cy}

\def\spa#1.#2{\left\langle#1\,#2\right\rangle}
\def\spb#1.#2{\left[#1\,#2\right]}

\def\tr{\mathop{\rm Tr}\nolimits}

\newbox\charbox
\newbox\slabox
\def\s#1{{      % Feynman slash
        \setbox\charbox=\hbox{$#1$}
        \setbox\slabox=\hbox{$/$}
        \dimen\charbox=\ht\slabox
        \advance\dimen\charbox by -\dp\slabox
        \advance\dimen\charbox by -\ht\charbox
        \advance\dimen\charbox by \dp\charbox
        \divide\dimen\charbox by 2
        \raise-\dimen\charbox\hbox to \wd\charbox{\hss/\hss}
        \llap{$#1$}
}}

\def\spa#1.#2{\langle#1\,#2\rangle}
\def\spb#1.#2{[#1\,#2]}
\def\lor#1.#2{\left(#1\,#2\right)}

\def\spba#1.#2.#3{[ #1  | K_{#2} | #3 \ra  }
\def\spaa#1.#2.#3.#4{\la #1 | K_{#2} K_{#3} | #4 \ra }
\def\s#1{s_{#1}}

\catcode`@=11  % Make @ letter.

\def\Tr{\, {\rm Tr}}

\def\la{\langle}
\def\ra{\rangle}

\def\lsl{\not{\hbox{\kern-2.3pt $\ell$}}}
\def\ksl{\not{\hbox{\kern-2.3pt $k$}}}

\def\tr{\mathop{\hbox{\rm Tr}}\nolimits}

\title{Identities amongst the Two Loop Partial Amplitudes of Yang-Mills theory}

\author[]{David~C.~Dunbar}

\affiliation[]{Department of Physics,\\
Faculty of Science and Engineering, \\
Swansea University,\\
Swansea, SA2 8PP, UK }

\emailAdd{d.c.dunbar@swansea.ac.uk}

\abstract{We characterise possible identities among the two-loop partial amplitudes of gluon scattering in Yang-Mills theory.  
We use known amplitudes in an exhaustive search to identify potential new relations.  We find two candidate relations which may extend
to all-$n$ amplitudes.}

\keywords{NNLO computations}

\begin{document}

\maketitle

%\TimeStamp
%\today
%%%%%%%%%%%%%%%%%%%%%%%%%%%%%%%%%%%%%%%%%%%%%%%%%%%%%%%%%%%%%%%%%%%%%
%%
%%
%%
%%  
%%
%%  
%%
%%
%%
%%%%%%%%%%%%%%%%%%%%%%%%%%%%%%%%%%%%%%%%%%%%%%%%%%%%%%%%%%%%%%%%%%%
\def\BB{B}
\def\BA{A}
\def\BC{C}
\def\TB{b}
\def\TA{a}
\def\TC{c}
\def\K{P}

\section{Introduction}

The scattering amplitudes of gluons within a pure $SU(N_c)$ gauge theory are  important from a phenomenological viewpoint  
where there is considerable demand for new predictions particularly 
at ``Next-to-Next-to-Leading Order'' (NNLO)~\citep{Amoroso:2020lgh}.  
Also,  amplitudes are the custodians of the symmetries 
of the theory  and as such are  important theoretical objects encapsulating information on the symmetries and properties of the theory.    

The amplitudes for gluon scattering are functions both of the kinematic variables of the scattered particles but also depend upon their gauge charges or color. 
Given this,  it is often convenient and informative to expand the full amplitude in terms of
color structures $C^\lambda$ multiplying partial amplitudes which contain the kinematic dependence
\begin{equation}
{\cal A}_n^{(\ell)} = \sum_{\lambda} { A}_{n:\lambda}^{(\ell)} C^\lambda\,,
\end{equation}
where $\ell$ denotes the loop order of the $n$-point amplitude.

An important expansion is where the color factors are products of the trace of color matrices.    
Gauge invariance implies that not all  the partial amplitudes are independent but that there may be relations amongst them possibly allowing less computational effort to determine the full color amplitude.    In particular ``decoupling identities'' which 
are obtained by extending the gauge group to $U(N_c)$ and looking at the relations necessary for the $U(1)$ gauge boson to
decouple.  However,  these do not exhaust the relations amongst partial amplitude and, for example, there exist Kleiss-Kuijf relations \citep{Kleiss:1988ne} among the tree amplitudes which 
coincide with the decoupling identities for low $n$ but 
are beyond decoupling identities in general.     Also, at one loop the subleading in color partial amplitudes can be expressed in terms of the leading~\citep{Bern:1994zx}: this identity again coinciding with decoupling for low $n$. 
In ref.~\citep{Naculich:2011ep} and \citep{Edison:2011ta,Edison:2012fn} it was shown that further identities among the two-loop amplitudes for four and five point amplitudes were obtained by using iterative methods assuming a three-point diagrammatic expansion.

In this paper we wish to explore the space of possible linear identities by examining the known amplitudes among two-loop amplitudes.   Specifically these are expressions of the form
\begin{equation}
\sum_\lambda a_\lambda { A}_{n:\lambda} ^{(2)}=0  \; , 
\label{eq:first}
\end{equation}
where the $a_\lambda$ are pure numbers not dependent upon the kinematic variables or helicity of the outgoing states.   

Unfortunately there are very few two-loop amplitudes in pure Yang-Mills for whom analytic forms are known which we can use to test and examine potential identities.  
Amplitudes are organised according to the helicity of the external gluon: we use the convention that all states are outgoing.   Only for 
the four point amplitude are all partial amplitudes known for all helicity states~\citep{Glover:2001af,Bern:2002tk}.    At five-point the leading in color amplitudes are known in analytic form from a series of papers.  Initially,  due to its simplicity,  the first calculation was for the special helicity configuration where all the (outgoing) helicities are positive (or all negative) which we describe as the ``all-plus'' amplitude~\citep{Gehrmann:2015bfy,Dunbar:2016aux}.   
These two loop amplitudes are considerably simpler in functional form than 
a general amplitude because the tree amplitude vanishes and consequently the one-loop amplitude is a pure rational function~\citep{Bern:1993qk}.    Subsequently the leading in color single minus amplitude~\citep{Badger:2018enw} and the remaining leading in color partial amplitudes~\citep{Abreu:2019odu} have been computed.  

Beyond the four-point amplitude and the leading in color at five point,  only amplitudes with the all-plus helicity configuration are know.   Specifically these are known for five  \citep{Badger:2019djh,Dunbar:2019fcq}
and six gluons.~\citep{Dalgleish:2020mof}.  Additionally a specific partial 
amplitude is known for $n$ points.~\citep{Dunbar:2020wdh}.

Using the all-plus five and six-point known amplitudes  we can verify the expected relations and determine if,  and how many,   further linear relations exist among the partial amplitudes.    Although relations identified for a particular helicity can only be conjectured to extend to all helicities we can definitely say there are no further identities beyond those satisfied by the all-plus amplitude.

\section{Color Decompositions}

In this section we will examine the color decomposition in detail.  We will start with a review of tree and one loop decompositions before discussing the two-loop amplitudes.  

In a Yang-Mills gauge theory with the gluons lying in the adjoint representation a $n$-gluon amplitude may be expanded in the gauge coupling constant,
\begin{equation}
{\cal A}_n = g^{n-2}  \sum_{\ell\geq 0} a^{\ell}{\cal A}_n^{(\ell)} 
\; , 
\end{equation}
where $a=g^2e^{-\gamma_E \epsilon}/(4\pi)^{2-\epsilon}$ and
${\cal A}_n^{(\ell)} $ is the $\ell$ loop amplitude which will be expanded in color structures as in eqn.~(\ref{eq:first}).

Our starting point is the color trace decomposition of a $n$-point gluon scattering amplitude at tree,  $\ell=0$, level within a $SU(N_c)$ or $U(N_c)$ gauge theory.  The full amplitude can be decomposed into 
partial amplitudes in a process which separates color and kinematics
\begin{eqnarray}
{\cal A}_n^{(0)}(1,2,3,\cdots ,n)  &=& \sum_{S_{n}/Z_n}  \Tr[ T^{a_1} T^{a_2} 
\cdots T^{a_n}] A_{n}^{(0)} (a_1,a_2,\cdots, a_n) , 
\label{eq:tree}\end{eqnarray}
where the amplitude is expressed in terms of the color matrices $T^{a_i}$ rather than the structure constants of the group. 
The partial amplitudes $A_{n}^{(0)} (a_1,a_2,\cdots, a_n)$ are cyclically symmetric but not fully crossing symmetric.
They are however fully gauge invariant since for sufficiently large $N_c$ the external gluons can have "color" choices which reduces the sum to a single term.   If,  we can use the cyclic symmetry to choose $a_1=1$, then  the sum over permutations is over the $(n-1)!$ permutations of $(2,\cdots, n)$.   Additionally,  the partial amplitudes are reflection symmetric 
\begin{eqnarray}
{A}_n^{(0)}(1,2,3,\cdots ,n)  &=& (-1)^n  A_{n}^{(0)} (n,\cdots, 3,2,1)
\; , 
\label{eq:treereflect}\end{eqnarray}
so there are $(n-1)!/2$  individual partial amplitudes.  The above decomposition can be derived from field theory
~\citep{Cvitanovic:1980bu,Mangano:1990by,Mangano:1988kk} but it  naturally arises in  string theory 
and consequently in field theory~\citep{Kosower:1987ic,Kosower:1988kh}.

Not all the partial amplitudes are independent but are related by gauge properties.  One route is to note that the expansion also applies to a $U(N_c)$ gauge group but must vanish if one (or more) of the external gluons carry a $U(1)$ charge.   Demanding this produces a {\it decoupling identity}.  
For example setting  leg $1$ to be $U(1)$ and extracting the coefficient of 
$\Tr[T^2 T^3 \cdots T^n]$ implies that
\begin{equation}
A_{n}^{(0)}(1,2,3,\cdots , n ) +A_{n}^{(0)}(2,1,3,\cdots  , n )+\cdots A_{n}^{(0)}(2,3, \cdots, 1, n )=0.
\label{eq:decotree}
\end{equation}

The expansion in terms of color traces is not unique and other expansions in terms of color structures exist.
In particular there is an expansion in terms of the structure constants~\citep{DelDuca:1999rs}
\begin{eqnarray}
{\cal A}_n^{(0)}(1,2,3,\cdots ,n)  &=& \sum_{S_{n-2}} 
f^{1a_2b_1}f^{b_1a_3b_2} \cdots f^{b_{n-1}a_{n-1}n}  A_{n}^{(0)} (1,a_2,\cdots, a_{n-1} ,  n) , 
\label{eq:decoDDM}
\end{eqnarray}
where we have selected legs $1$ and $n$ and the summation is over the $(n-2)!$ permutations $(a_2,a_3,\cdots, a_{n-1})$ of the remaining legs $2,\cdots , n-1$.

The partial amplitudes in eqn.~\eqref{eq:decoDDM} are the same as in 
eqn.~\eqref{eq:tree}.     To equate the two expressions and to show this requires the following identity among the partial amplitudes, 
\begin{equation}
A_{n}^{(0)} (1,\{\alpha\}, n, \{\beta\}) =  
(-1)^{|\beta|} \sum_{\sigma \in OP(\alpha ,\beta^T)}  
A_{n}^{(0)}( 1 , \{\sigma\} , n)
\label{eq:KK}
\end{equation}
where $\alpha$ and $\beta$ are some sets of the remaining indices i.e. 
$\{\alpha\} =\{ a_2 , \cdots a_p\}$ and $\{\beta\} =\{a_{p+1},\cdots a_{n-1}\}$.   
The summation is over the order permutations of $\alpha$ and $\beta^T$. That is permutations of the union of the sets where the ordering of 
$\alpha$ and $\beta^T$ are preserved.   The summation contains $(r+s)!/r!s!$ terms where $r,s$ are the number of indices in the sets $\alpha$ and $\beta$ respectively.  In appendix~\ref{app:permutations},  specific examples of 
order permutations are given. 
 
These identities, known as Kleiss-Kuijf relations~\citep{Kleiss:1988ne},  overlap with the decoupling identities but contain for $n \geq 7$ more information.   For $n \leq 6$ decoupling identities are sufficient to prove eqn.~\eqref{eq:KK} but for $n>6$ the rank of the 
system of decoupling identities in not large enough to prove eqn.~\eqref{eq:KK} (although for $n=7$  the rank is {\it just} insufficient).    
We can see this from table~1  where we have evaluated the number of independent conditions (rank) obtained by all possible decoupling equations.   For $n=6$ this rank of 36 is exactly that needed to express the 
60 $A_6^{(0)}(1,\sigma)$ in terms of the 24 $A_{6}^{(0)}(1,\sigma,n)$. However for $n=7$ there is (just) insufficient information with the disparity increasing with $n$.    In practice,  at $n=7$ we cannot find decoupling relations which express
\begin{equation}
A^{(0)}_{7} ( 1, a,b,c , 7, d,e )
\end{equation} 
in terms of $A_{7}^{(0)}(1,\sigma,7)$ although the combination   \begin{equation}
A^{(0)}_{7} ( 1, a,b,c , 7, d,e )+A^{(0)}_{7} ( 1, a,b,7,c , d,e ) 
\end{equation}
may be. 

 Beyond $n=7$,  it is the
$A^{(0)}_{n} ( 1,  S_1,  n ,  S_2 )$ where the length of both $S_i$ is at least two and one has length greater than two which the decoupling identities fail to determine.

The partial amplitudes $A_{n}^{(0)} (1,a_2,\cdots, a_{n-1} ,  n)$ form a minimal set in that there are no further linear relations between them valid for all helicities as can be verified by examining the known tree amplitudes. 

\begin{table}[h] 
\begin{tabular}{c|c|c|c|c}
\hline\hline
n & $\#$ of $A_{n}^{(0)}(1,\sigma)$ & Rank of Decoupling System  & $\#$ of $A_{n}^{(0)}(1,\sigma,n)$ &  $\#$ of extra relations  \\
\hline
6 & 60  & 36   & 24 &  0\\
7 & 360 & 239  & 120 & 1 \\
8 & 2520 &   1696    & 720 & 104  
\\
\end{tabular}
\label{DecoTable}
\caption{This table enumerates the rank of the system of decoupling identities and demonstrates this is not enough to  determine eqn.~\eqref{eq:KK} for $n> 6$. The final column shows the number of extra relations contained in the Kleiss-Kuijf relations beyond the decoupling identities.}
\end{table}

We are interested in this article in relations between amplitudes with constant coefficients. For tree amplitudes there are further relations if we allow the coefficients which depend upon kinematic factors but which are independent of helicity.  These rely upon the Bern, Carrasco and Johansson relation (BCJ) for gauge theories~\citep{Bern:2008qj}.  These can reduce the $(n-2)!$  basis to a $(n-3)!$ basis for tree amplitudes~\citep{Bjerrum-Bohr:2009ulz}.

At one-loop the decomposition in terms of color traces contains both single trace terms 
and double trace terms~\citep{Bern:1990ux}
\begin{eqnarray}
{\cal A}_n^{(1)}(1,2,&3& ,\cdots ,n)   = \sum_{S_n/Z_n}  N_c  \Tr[ T^{a_1} T^{a_2} 
\cdots T^{a_n}] A_{n:1}^{(1)} (a_1,a_2,\cdots, a_n)
\notag
\\
+ \sum_{r}   \sum_{P_{n:r}}   &&\hskip -15pt
\Tr[ T^{a_1} 
\cdots T^{a_{r-1}}]  \Tr[ T^{a_{r}} 
\cdots T^{a_n}] A_{n:r}^{(1)} (a_1,a_2,\cdots, a_{r-1} ; a_{r}, \cdots ,a_n) \; .
\end{eqnarray}
The single trace term has a factor of $N_c$ and is thus referred to as the leading-in-color (or planar) contribution. 
The summation over $r$ for a $SU(N_c)$ theory is $r=3,\cdots [n/2]$. For a $U(N_c)$ theory the summation is over 
$r=2,\cdots [n/2]$ with the $r=2$  term $A_{n:2}^{(1)} (a_1 ; a_2,\cdots, a_n)$ which has color structure 
$\Tr(T^{a_1})\Tr(T^{a_2} \cdots T^{a_n})$.   We refer to partial amplitudes such as $A_{n:2}^{(1)}$ as a $U(N_c)$ specific amplitude.

For the one-loop amplitude the double trace terms are not independent but
can be expressed in terms of the leading
\begin{equation}
A_{n:r}^{(1)} (a_1,a_2,\cdots, a_{r-1} ; a_{r}, \cdots , a_n)=(-1)^{r} \sum_{\sigma\in COP\{\alpha\}\{\beta^T\}} 
A_{n:1}^{(1)}  (\sigma )
\label{eq:OneLoopSubLeading}
\end{equation}
where $\alpha=\{ a_1 ,\cdots a_{r-1}\}$ and $\beta=\{a_{r} \cdots a_n \}$.   The summation is over the ordered permutations as before but factoring out equivalent permutations due to cyclic symmetry (see appendix~\ref{app:permutations} for examples). 
This relation allows the double trace terms to be derived from the leading in color terms only.  This reduces the number of functional forms to be computed in a calculation considerably.

 There are some analogues between the one-loop relation eqn.~(\ref{eq:OneLoopSubLeading}) and the tree relation eqn.~(\ref{eq:KK}).  
 This relation can be obtained from the decoupling equations for $n\leq 5$ but beyond $n=5$ the decoupling equations are not sufficient: e.g.  see \citep{Feng:2011fja}. Explicitly,  at $n=6$,   decoupling identities determine the combination 
 \begin{equation}
 A_{6:4}^{(1)}  (a_1,a_2,a_3  ;  a_4,a_5,a_6 )+A_{6:4}^{(1)}  (a_1,a_3,a_2  ;  a_4,a_5,a_6 )
 \end{equation}
 but in themselves do not determine the individual terms. 

The relation (\ref{eq:OneLoopSubLeading}) can be shown in multiple ways.   It is a natural relation when Yang-Mills theory is viewed as the low energy limit of open string theory~\citep{Bern:1994zx}.    In open string theory,  the gauge content of a $U(N_c)$ theory is carried by the Chan-Paton factors at the string ends which carry the color charge of quarks and anti-quarks respectively.    The amplitude is given by an integration over all possible world sheets.  External adjoint states are obtained by inserting a vertex operator with corresponding color matrix to a boundary.  The equivalent for one-loop is a world sheet with two boundaries.  The contribution to a term
\begin{equation}
\Tr[ T^{a_1} \cdots T^{a_{r-1}}]  \Tr[ T^{a_{r}} 
\cdots T^{a_n}] 
\end{equation}
arise from where states $a_1$ to $a_{r-1}$ are attached to a single boundary and states $a_r$ to $a_n$ are attached to the other.    In ref~\citep{Bern:1994zx}  this diagrammatic view was developed into a proof of the relation (\ref{eq:OneLoopSubLeading}).   In ref~\citep{Feng:2011fja} the relation was derived using  unitarity methods~\citep{Bern:1994zx,Bern:1994cg}   together with the Kleiss-Kuif relations.

Note that the leading in color terms do not satisfy the Kleis-Kuif relation~\eqref{eq:KK} in themselves.  The simplest decoupling identity being
\begin{equation}
A_{n:2}^{(1)}(1 ; 2,3,\cdots , n ) +  A^{(1)}_{n:1}(1,2,3,\cdots , n ) 
+A_{n:1}^{(1)}(2,1,3,\cdots  , n )+\cdots A_{n:1}^{(1)}(2,\cdots, 1, n )=0.
\label{eq:decoloop}
\end{equation}
This is similar to the tree decoupling but with the additional 
$A_{n:2}^{(2)}$ term.  The decoupling identities will then imply (at least for $n\leq 6$)
\begin{equation}
A_{n:1}^{(1)} (1,\{\alpha\}, n, \{\beta\}) =  
(-1)^{|\beta|} \sum_{\sigma \in OP(\alpha ,\beta^T)}  
A_{n:1}^{(1)}( 1 , \{\sigma\} , n) +\sum A_{n:2}^{(1)}
\label{eq:KKloop}
\end{equation}
so unless the functions $A_{n:2}^{(1)}$ are zero eqn.~\eqref{eq:KK} is not  satisfied by the $A_{n:1}^{(1)}$. 
\def\Explicitly{Explicitly for $n=6$, 
\begin{eqnarray}
A_{6:1}^{(1)} (1,\{\alpha_1,\alpha_2,\alpha_3\}, 6, \{\beta_1\}) &=&  
- \sum_{\sigma \in OP(\alpha ,\beta^T)}  
A_{6:1}^{(1)}( 1 , \{\sigma\} , 6) -A_{6:2}^{(1)} (\beta_1 ; 1,\alpha_1,\alpha_2,\alpha_3,6)
\notag
\\
A_{6:1}^{(1)} (1,\{\alpha_1,\alpha_2\}, 6, \{\beta_1,\beta_2 \}) &=&  
 \sum_{\sigma \in OP(\alpha ,\beta^T)}  
A_{6:1}^{(1)}( 1 , \{\sigma\} , 6) 
\label{eq:KKloop6}\\
-A_{6:2}^{(1)} (6  ; 1,\alpha_1,\alpha_2,\beta_1,\beta_2)
&+& A_{6:2}^{(1)} (\alpha_1 ; 1,6, \alpha_2,\beta_1,\beta_2)
+A_{6:2}^{(1)} (\beta_2 ; 1, 6, \alpha_1,\alpha_2,\beta_1)
\notag
\end{eqnarray}
Note these imply identities of the form $
\sum a_i  A_{n:2}^{(1)}=0$.    }

The explicit form for $A_{n:2}^{(1)}$ of the all-plus helicity amplitude is
\begin{eqnarray}
A_{n:2}^{(1)}(1^+ ;2^+,3^+,\cdots, n^+) &=& -i\frac{1}{\spa2.3\spa3.4 \cdots \spa{n}.2} \sum_{2 \leq i<j \leq n}  \spb{1}.{i} \spa{i}.{j} \spb{j}.{1} 
\label{eq:oneloopsubleading}
\end{eqnarray}
which is explicitly non-vanishing in eqn.~(\ref{eq:KKloop}). 
This expression is from reference~\citep{Dunbar:2019fcq}.   
Although we can regard the  $U(N_c)$ specific amplitudes as un-physical they are gauge invariant and as such appear as building blocks in other amplitudes -this particular partial amplitude is also the one-loop amplitude between a single photon and $n-1$ gluons (all of positive helicity)  due to a scalar or fermion loop for which an earlier explicit analytic form exists~\citep{Bern:1993qk}.   The leading in color all-plus amplitude is related to the $N=4$ MHV amplitude by a dimension shifting of the 
loop integrals~\citep{Bern:1996ja,Britto:2020crg}.
 Note that the expression in eqn.~(\ref{eq:oneloopsubleading})  matches that obtained from eqn.~(\ref{eq:OneLoopSubLeading}) but only after significant simplification.    Analytic expressions derived from the decoupling and other such identities are often
inefficient.

We now turn to the topic of this letter. 
A general two-loop amplitude for the scattering of $n$ gluons in a pure $SU(N_c)$ or $U(N_c)$ gauge theory 
may be expanded in a color trace basis as
\begin{eqnarray}
& & {\cal A}_n^{(2)}(1,2,\cdots ,n) =
N_c^2 \sum_{S_n/\mathcal{P}_{n:1}}  \tr[T^{a_1}T^{a_2}\cdots T^{a_n}] A_{n:1}^{(2)}(a_1,a_2,\cdots ,a_n) \notag \\
&+&
N_c\sum_{r=2}^{[n/2]+1}\sum_{S_n/\mathcal{P}_{n:r} }   \tr[T^{a_1}T^{a_2}\cdots T^{a_{r-1}}]\tr[T^{b_r} \cdots T^{b_n}] 
A_{n:r}^{(2)}(a_1,a_2,\cdots ,a_{r-1} ; b_{r}, \cdots, b_n)  
\notag \\
&+& \sum_{s=1}^{[n/3]} \sum_{t=s}^{[(n-s)/2]}\sum_{S_n/\mathcal{P}_{n:s,t}} 
%\hskip -1.0 truecm 
\tr[T^{a_1}\cdots T^{a_s}]\tr[T^{b_{s+1}} \cdots T^{b_{s+t}}]
\tr[T^{c_{s+t+1}}\cdots T^{c_n}] 
\notag
\\
& & 
\hskip 7.0truecm 
\times A_{n:s,t}^{(2)}(a_1,\cdots ,a_s;b_{s+1} ,\cdots, b_{s+t} ;c_{s+t+1},\cdots, c_n ) 
\notag \\
&+&\sum_{S_n/\mathcal{P}_{n:1}}  \tr[T^{a_1}T^{a_2}\cdots T^{a_n}] A_{n:1B }^{(2)}(a_1,a_2,\cdots ,a_n)\,.
\label{eq:twoloopexpansion}
\end{eqnarray}
The partial amplitudes multiplying any trace of color matrices are cyclically symmetric in the indices within the trace.  
The expression has single, double and triple trace terms.  In string theory these would arise from surfaces with three boundaries.  
The is also a single trace term $A_{n:1B }^{(2)}$ which is sub-sub leading in powers of $N_c$.  This arises in string theory from a separate two-loop surface with a single boundary~\citep{Dunbar:2020wdh}.

The summations simply count each color structure exactly once.    Specifically, 
when the sets are of different lengths ($r-1\neq \frac{n}2$, $s\neq t $, $t\neq\frac{n-s}{2}$ and $3s\neq m,n$)
the sets $\mathcal{P}_{n:\lambda}$ 
are
\begin{align}
\mathcal{P}_{n:1}&=Z_{n}(a_1,\cdots, a_n),\notag\\
\mathcal{P}_{n:r}&=Z_{r-1}(a_1,\cdots,a_{r-1})\times Z_{n+1-r}(a_r,\cdots,a_n),  \;\;\ r > 1, r-1 \neq n+1-r  \notag\\
\mathcal{P}_{n:s,t}&=Z_s(a_1,\cdots,a_s)\times Z_t(a_{s+1},\cdots,a_{s+t})\times Z_{n-s-t}(a_{s+t+1},\cdots,a_n)\,.
%\label{eq:manifestsets}
\end{align}
When the sets have equal lengths, to avoid double counting
\begin{align}
\mathcal{P}_{2m:m+1}&=Z_{m}(a_1,\cdots,a_m)\times Z_{m}(a_{m+1},\cdots,a_{2m})\times Z_2, \\
\mathcal{P}_{n:s,s}&= Z_s(a_1,\cdots,a_s)\times Z_s(a_{s+1},\cdots,a_{2s})\times Z_{n-2s}(a_{2s+1},\cdots,a_n)\times Z_2,\notag\\
\mathcal{P}_{3m:m,m}&=Z_{m}(a_1,\cdots,a_m)\times Z_{m}(a_{m+1},\cdots,a_{2m})\times Z_{m}(a_{2m+1},\cdots,a_{3m})\times S_{3},\notag \\
\mathcal{P}_{2m:2s,m-s}&= Z_{2s}(a_1,\cdots,a_{2s})\times Z_{m-s}(a_{2s+1},\cdots,a_{s+m})\times 
Z_{m-s}(a_{s+m+1},\cdots,a_{2m})\times Z_2  \,.\notag
%\label{eq:manifestsets}
\end{align}
The partial amplitudes are
\begin{equation}
A_{n:r}^{(2)}  \;\;  r=1 \cdots [n/2]   \; \;\;\  A_{n:r,s}^{(2)}    \;\;  r,s =1 \cdots [n/3],  r \leq s
\end{equation}
of which the $A_{n:2}^{(2)}$ and $A_{n:1, s}^{(2)}$ are the $U(N_c)$ specific functions.

The two-loop expansion is  an expansion in powers of $N_c$.  Decoupling identities do involve different powers of $N_c$ since if we set 
\begin{equation}
T^1 \longrightarrow T^{U(1)} = \frac{1}{\sqrt{N_c}} I
\end{equation}
then 
\begin{equation}
N_c^2 \tr \left[ T^1 T^{a_2} \cdots T^{a_n} \right] \longrightarrow N_c^{3/2} \tr  \left[T^{a_2} \cdots T^{a_n} \right]
\end{equation}
and 
\begin{equation}
N_c^{1}  \tr \left[ T^1\right] \tr  \left[   T^{a_2} \cdots T^{a_n} \right] \longrightarrow N_c^{3/2} \left[ T^{a_2} \cdots T^{a_n} \right]
\end{equation}
and equating the coefficient of $ \tr  \left[T^{a_2} \cdots T^{a_n} \right]$ gives a decoupling identity amongst terms of different order in $N_c$.     However,  decoupling identities do not  relate the single trace amplitudes $A_{n:1B}^{(2)}$ to the other amplitudes. Instead they obey decoupling identities among themselves identical to those satisfied by the tree amplitudes 
$A_{n}^{(0)}$.   Consequently,   for $n \leq 6 $ these are guaranteed to obey the same relation shown in eqn.~(\ref{eq:KK})

Decoupling identities can be used to express the $U(N_c)$ specific amplitudes in terms of the 
In terms of $SU(N_c)$ partial amplitudes,.  For example, from the above we have
\begin{equation}
A_{n:2}^{(2)}(1 ; 2,3,\cdots,n)  +A_{n:1}^{(2)}(1,2,3,\cdots, n ) +A_{n:1}^{(2)}(2,1,3,\cdots, n )+\cdots +A_{n:1}^{(2)}(2,\cdots, 1, n )  =0 \,.
\end{equation}
This allows $A_{n:2}^{(2)}$ to be expressed  in terms of the 
$A_{n:1}^{(2)}$.     
For the $SU(N_c)$ amplitudes
decoupling identities are relatively limited only determining the triple trace partial amplitudes $A_{n:2,r}^{(2)}$.

\noindent
In total,  after setting $T^1$ to be a $U(1)$,

$\bullet$ the coefficient of $\Tr (T^2 \cdots T^n)$ fixes $A_{n:2}^{(2)}$ as a sum of 
$A_{n:1}^{(2)}$  amplitudes

$\bullet$ the coefficient of $\Tr (T^2 \cdots T^{s-1} )\Tr(T^s \cdots T^n)$ fixes 
 $A_{n:1, s}^{(2)}$  is terms of $A_{n:s}^{(2)}$ and $A_{n:s+1}^{(2)}$

$\bullet$   the coefficient of  
%\begin{equation}
$\Tr(T^2)\Tr(T^3\cdots T^{s+2})\Tr( T^{s+3} \cdots T^n)$
%\end{equation}
enables the $SU(N_c)$ functions the $A_{n:2,s}^{(2)}$ to be fixed
as
\begin{eqnarray}
A_{n:2, s}^{(2)} (1,2\;  ;  3 \cdots s+3 ; s+4 \cdots n )
=& & -\sum  A_{n:1, s}^{(2)}  (2 ;1, 3, \cdots s+3 ; s+4 \cdots n )
\notag
\\
 -\ \sum & & A_{n:1, s+1}^{(2)}  (2 ;  3 \cdots s+3 ; 1, s+4 \cdots n )
%\notag
\\ = -\hskip -1.0truecm\sum_{\sigma \in COP( \{3,\cdots s+3\},\{1\})}  A_{n:1, s}^{(2)}  (2 ; \sigma ; s+4, \cdots n)
& &-\hskip -25pt \sum_{\sigma \in COP( \{s+4,\cdots n\},\{1\})}   A_{n:1, s}^{(2)}  (2 ; 3,\cdots s+3 ; \sigma )  
\notag
\label{eq:An2r}\end{eqnarray}
where the summation denotes summing over the different locations leg one may appear. 
This expression is likely to be the simplest but the $U(N_c)$ specific functions can be substituted to leave an expression 
purely in terms of the $SU(N_c)$ functions.

The first triple trace term which cannot be determined from decoupling identities will be the nine-point partial amplitude $A_{9:3,3}^{(2)}$ for which decoupling identities only determine the combination
\begin{equation}
A_{9:3,3}^{(2)}(1,2,3 ; 4,5,6 ; 7,8,9) +A_{9:3,3}^{(2)}(1,3,2 ; 4,5,6 ; 7,8,9) 
\end{equation}

In the following sections we review the known identities for $n=5$ and review and explore possible new identities for $n=6$.

\section{Identities among Five point partial amplitudes}

In this section we will review the known identities amongst the five point partial amplitudes. 

The five point partial amplitudes in the color-trace basis expansion are
\begin{equation}
A_{5:1}^{(2)}  , A_{5:2}^{(2)}  ,  A_{5:3}^{(2)} ,   A_{5:1,1}^{(2)} ,   A_{5:1,2}^{(2)}   \hbox{ and }  A_{5:1B}^{(2)} 
\end{equation}
of which the three 
\begin{equation}
A_{5:1}^{(2)} \; (12)  ,  A_{5:3}^{(2)} \; (15) \hbox{ and }  A_{5:1B}^{(2)} ; (12)
\end{equation}
are the $SU(N_c)$ functions.   The numbers in brackets indicate the number of independent amplitudes of each type. 
The $A_{5:1,2}^{(2)}$ vanish since $ A_{5:1,2}^{(2)}(a;b,c;d,e)=-A_{5:1,2}^{(2)}(a;c,b;e,d)=-A_{5:1,2}^{(2)}(a;b,c;d,e)$.

We now review the known identities among these partial amplitude. 
Firstly,  taking the decoupling identities among the $A_{5:1B}^{(2)}$.  For these the decoupling identity system has rank 6 which allows a Kleiss-Kuijf relation leaving the six $A_{5:1B}^{(2)}(1,a,b,c,5)$ as independent functions.   
For the remaining functions the decoupling identities only determine the form of the specifically $U(N_c)$ functions 
$A_{5:2}^{(2)}$ and  $A_{5:1,1}^{(2)}$ in terms of $A_{5:3}^{(2)}$ and $A_{5:1}^{(2)}$ but place no further constraints upon these functions.

In ref~\citep{Edison:2011ta} using iteration and color kinematic duality a further 6 relations were found.  Unlike decoupling identities these involved both the $A_{5:1B}^{(2)}$  and other partial amplitudes and indeed allowed for a solution  for $A_{5:1B}^{(2)}$ in terms of $A_{5:1}^{(2)}$  and  $A_{5:3}^{(2)}$.  
\begin{eqnarray}
A_{5:1B}^{(2)}(1,2,3,4,5)   &=&  
-A_{5:1}^{(2)}(1, 2, 4, 3,  5) 
+2 A_{5:1}^{(2)}(1, 2, 5, 3, 4) 
+A_{5:1}^{(2)}( 1, 2, 5, 4, 3)
\notag \\
& &-A_{5:1}^{(2)}( 1, 3, 2, 4, 5) 
+2 A_{5:1}^{(2)}( 1, 3, 4, 2, 5) 
-5 A_{5:1}^{(2)}( 1, 3, 5, 2, 4) 
\notag \\
& &-2 A_{5:1}^{(2)}( 1, 3, 5, 4, 2 )  
+2 A_{5:1}^{(2)}( 1, 4, 2, 3, 5 ) 
+A_{5:1}^{(2)}( 1, 4, 3, 2, 5 ) 
\notag \\
& &+2 A_{5:1}^{(2)}(1, 4, 5, 2, 3)
+A_{5:1}^{(2)}(1, 4, 5, 3, 2 )
\notag \\
&-&\frac{1}{2} \sum_{Z_5(1,2,3,4,5)} \bigg(   A_{5:3}^{(2)}( 1 ,2 ;  3,4,5 ) 
 -A_{5:3}^{(2)}( 1,3 ; 2,4,5 ) \biggr)\,.
\label{eq:Edi}
\end{eqnarray}
which reduces the number of functional forms necessary to calculate from three to two.  

This expression is rather strange in several respects.   Firstly,  it contains factors which are  a rather unnatural and secondly the {\it only}   leading in color term $A_{5:1}^{(2)}$ missing is $A_{5:1}^{(2)}(1,2,3,4,5)$. I.e. the only term missing is that with the same ordering of legs as the $A_{5:1B}^{(2)}$.   
Part of the structure is driven by symmetry.  Assuming that 
$A^{(2)}_{5:1B}$ can be expressed in terms of $A^{(2)}_{5:1}$ and 
$A^{(2)}_{5:3}$ and demanding the reflection and cyclic symmetries of 
$A_{5:1B}^{(2)}$ would imply
\begin{eqnarray}
A_{5:1B}^{(2)}(1,2,3,4,5)   &=&  
a_1(   A_{5:1}^{(2)}(1, 2, 3, 4,  5) )
+a_2 ( A_{5:1}^{(2)}( 1, 3, 5, 2, 4) ) 
\notag \\
+a_3(  
A_{5:1}^{(2)}(1, 2, 4, 3,  5) 
-A_{5:1}^{(2)}( 1, 2, 5, 4, 3)
& & \hskip -0.5 truecm
+A_{5:1}^{(2)}( 1, 3, 2, 4, 5) 
-A_{5:1}^{(2)}( 1, 4, 3, 2, 5 ) 
-A_{5:1}^{(2)}(1, 4, 5, 3, 2 )
)
\notag \\    \hskip -0.5 truecm
+a_4 (A_{5:1}^{(2)}(1, 2, 5, 3, 4) 
+A_{5:1}^{(2)}( 1, 3, 4, 2, 5 )  
&&\hskip -0.5 truecm
-A_{5:1}^{(2)}( 1, 3 ,5, 4, 2 ) 
+A_{5:1}^{(2)}(1, 4, 2, 3, 5)
+A_{5:1}^{(2)}( 1, 4, 5, 2, 3)   )
\notag \\
+a_5 \sum_{Z_5(1,2,3,4,5)}    A_{5:3}^{(2)}( 1 ,2 ;  3,4,5 ) 
&&+a_6  \sum_{Z_5(1,2,3,4,5)} A_{5:3}^{(2)}( 1,3 ; 2,4,5 )
\label{eq:EdiSym}
\end{eqnarray}
which matches the correct term with $a_1=0$, $a_5=-5$ etc.

The expression has been written in terms of the $SU(N_c)$ amplitudes $A_{5:1}^{(2)}$ and $A_{5:3}^{(2)}.$ It can be rewritten in  terms of the $U(N_c)$ specific amplitudes in several alternate forms:
\begin{eqnarray}
A_{5:1B}^{(2)}(1,2,3,4,5)   &=&
\sum_{Z_5(1,2,3,4,5)} \bigg(   
A_{5:2}^{(2)}( 1 ;3 , 5,2,4 ) 
-\frac{1}{2} A_{5:3}^{(2)}( 1 ,2 ;  3,4,5 ) 
 +\frac{1}{2} A_{5:3}^{(2)}( 1,3 ; 2,4,5 ) \biggr)\,.
\notag 
\\
\hbox{ or} \notag 
\\
A_{5:1B}^{(2)}(1,2,3,4,5)   &=&  \frac{1}{2} \sum_{Z_5(1,2,3,4,5)} \bigg(  
A_{5:1, 1}^{(2)}( 1 ,2 ;  3,4,5 ) -A_{5:1, 1}^{(2)}( 1 ,3 ;  2,4,5 )  \bigg)
\end{eqnarray}
which has more natural factors. 

Part of the purpose of this article will be an exhaustive search for possible relation like~(\ref{eq:Edi}) for $n >5$.

\section{Identities among Six point partial amplitudes}
In the expansion of the two loop amplitude~(\ref{eq:twoloopexpansion})
at six-point there are the following partial amplitudes
\begin{equation}
A_{6:1}^{(2)}  , A_{6:2}^{(2)}  ,  A_{6,3}^{(2)} ,   A_{6:4}^{(2)}  ,   A_{6:1,1}^{(2)} ,   A_{6:1,2}^{(2)}  , A_{6:2,2}^{(2)}  \hbox{ and }  A_{6:1B}^{(2)} 
\end{equation}
Of these the $SU(N_c)$ amplitudes are
\begin{equation}
A_{6:1}^{(2)} \; (60)  ,  A_{6:3}^{(2)} \;(45)  ,   A_{6:4}^{(2)} \; (20)  ,   A_{6:2,2}^{(2)} \; (15) \;   \hbox{ and  } A_{6:1B}^{(2)} \;(60)
\end{equation}
with the number of independent amplitudes after applying cyclic and reflection symmetry given in brackets. 
Decoupling identities imply that the $A_{6:2,2}^{(2)} $ can be expressed in terms of the others via~(\ref{eq:An2r})  leaving four
partial amplitudes which need to be computed for an $SU(N_c)$ amplitude.

We will be exploring possible linear identities between these partial amplitudes. These will fit into three groups

1.  Those obtained by decoupling identities.  This system of identities has rank 207.  Once the $U(N_c)$ specific partial amplitudes have been solved for the system of decoupling identities has rank sixty-six (66).  There split into two distinct groups.  Thirty-six (36) exclusively involve $A^{(2)}_{6:1B}$ whilst the other thirty (30) involve purely the remaining $SU(N_c)$ amplitudes.    The thirty six identities can be used to reduce the $A^{(2)}_{6:1B}$ amplitudes to 
the independent 24 $A^{(2)}_{6:1B}(1,a,b,c,d,6)$.

2. Those obtained by iterative methods as Edison and Naculich not in 1. (14).  These are different from the decoupling identities in that they involve both the $A^{(2)}_{6:1B}$ and the others.     However numerically there are not enough to solve for the 
24 $A^{(2)}_{6:1B}$ .  

3. Identities  which are satisfied by the partial amplitudes of the two-loop all-plus amplitude.
It is the third type we will be looking for.   In particular,   we will investigate whether the number of necessary partial amplitudes may be reduced from four.

Identities of types 1. and 2.  are guaranteed to be satisfied for all helicities.  Identities of type 3 are not so guaranteed however they  set an envelope on possible all-helicity identities

First we examine the information given by the decoupling identities after solving for the $U(N_c)$ specific functions. 
For six points,  among the $SU(N_c)$ the remaining decoupling identities form a system of rank 30.   This determines the 
partial amplitude $A^{(2)}_{6:2,2}$ via
 \begin{eqnarray}
A^{(2)}_{6:2,2} (a,b  ; r,s ; m,n)
&=&A_{6:3}^{sym}(r,s ; a,b,m,n)
+A_{6:3}^{sym}(m,n ; a,b,r,s)
\notag \\
&+&A_{6:4}^{sym}(a, m,n ; b,  r,  s)
+A_{6:4}^{sym}(b,m,n ; a,  r,  s)
\label{eq:A622first }\end{eqnarray}
where we define the combinations
\begin{equation}
A_{6:4}^{sym}(a,b,c ;  r,  s,  t) \equiv  A_{6:4}^{(2)}(a,  b,  c ;  r,  s,  t) + A_{6:4}^{(2)}(a,  b,  c ;  r,  t,  s) 
\end{equation}
and
\begin{equation}
A_{6:3}^{sym}  ( a,b ; r,s,t,u) \equiv
A^{(2)}_{6:3} ( a,b ; r,s,t,u )+
A^{(2)}_{6:3} ( a,b ; r, t,s ,u )+
A^{(2)}_{6:3}  ( a,b ; r,s,u,t )
\end{equation}
The decoupling identities can be used to solve for the combination $A_{6:4}^{sym}$ with
\begin{equation}
A_{6:4}^{sym}(a,b,c ; r,s,t ) = \frac{1}{2} \sum_{\sigma}  A_{6:1}^{(2)}(\sigma)
-\frac{1}{2}  \left( A^{sym}_{6:3}(a,  b ; c,  r,  s,  t)+  A^{sym}_{6:3}(a,c ; b,  r,  s,  t)+  A^{sym}_{6:3}( b,c ; a,  r,  s,  t) \right)
\label{eq:DecCona}
\end{equation}
where the summation is over the entire 60 $A_{6:1}^{(2)}$.   This amounts to 10 independent constraints.  
Note this expression is not symmetric between the first and second triple of indices so we obtain the identities 
among the $A_{6:3}^{(2)}$
\begin{eqnarray}
A^{sym}_{6:3}(a,  b  ;  c,  r,  s,  t)+  
A^{sym}_{6:3}( b,  c  ;  a,  r,  s,  t)+  
A^{sym}_{6:3}(c,  a,  ;  b,  r,  s,  t)
\notag\\
=
A^{sym}_{6:3}(r,   s   ;  t,  a,  b,  c)+
 A^{sym}_{6:3}(s,    t  ;  r,  a,  b,  c)+
 A^{sym}_{6:3}(t,    r   ;  s  a,  b,  c)
\label{eq:DecConb}
\end{eqnarray}
This set of (10) identities has rank 5 so eqn.~(\ref{eq:DecCona}) and eqn.~(\ref{eq:DecConb}) are the final consistency constraints among the partial amplitudes determined by the decoupling identities and as such are satisfied for all helicity configurations. 

We can now examine possible identities in the amplitudes beyond these.   We expect at least 14 from the results
of ref.~\citep{Edison:2012fn}.    Our methodology is to consider relations, 
\begin{equation}
\sum  c_{\lambda,i}   R^{(2)}_{6:\lambda}(\sigma_i)  =0
\end{equation}
where the $R^{(2)}_{6:\lambda}$ are the rational parts of the six-point all-plus amplitude and to evaluate that as sufficient kinematic points with rational momenta that the system can be solved analytically for rational constants $c_{\lambda ,i}$.     We again comment that this method will identify any possible identities but these are only potential identities which may or may not extend to all amplitudes.  We use the rational terms calculated in ref.~\citep{Dunbar:2019fcq} which are include in appendix~\ref{app:sixpt}.
The rational terms are only part of the amplitude.   After identifying possible relations we check whether these also apply to the
polylogarithmic parts- these are also available in  ref.~\citep{Dunbar:2019fcq}.  

In total,  we find a further 20 identities among the rational terms of partial amplitudes.  I.e.  6 beyond that of~\citep{Edison:2012fn}.      An obvious initial comment is that these will {\it not} be enough to solve for the 24 $A_{6:1B}^{(2)}$.  In fact, they split into 14 identities involving $A_{6:1B}^{(2)}$ plus six which only involve the others.  
Before looking at the new relations we briefly comment upon the 14 involving $A_{6:1B}^{(2)}$.

  In particular,   if we consider the 24 $A_{6:1B}^{(2)}$ 
then  the rank of the identities involving these is only 14  as recognised in identified in 
ref~\citep{Edison:2012fn} and confirmed by our studies of the rational parts of the all-plus so it is not possible to solve for these. Instead the possible identities involve the combinations
\begin{eqnarray} 
A_{6:1B}^{sym_1}
&\equiv&  
\sum_{sym(b,c,d)} A_{6:1B}^{(2)} (1,a,b,c,d,6 )
\notag\\
 A_{6:1B}^{sym_2} 
&\equiv&  
\sum_{sym(a,b,c)}  A_{6:1B}^{(2)} (1,a,b,c,d,6 )
\notag\\
A_{6:1B}^{sym_3}
&\equiv &  
A_{6:1B}^{(2)} (1,a,b,c,d,6 )+
 A_{6:1B}^{(2)} (1,a,c,b,d,6 )\
\notag \\ & & +
 A_{6:1B}^{(2)} (1,d,b,c,a,6 )+
 A_{6:1B}^{(2)} (1,d,c,b,a,6 ) 
 \notag\\ 
 A_{6:1B}^{sym_4}
&\equiv&  
 A_{6:1B}^{(2)} (1,a,b,c,d,6 )+ A_{6:1B}^{(2)} (1,a,b,d,c,6 )
 +A_{6:1B}^{(2)} (1,c,a,b,d,6 )
\notag \\ &+& A_{6:1B}^{(2)} (1,d,a,b,c,6 )
 + A_{6:1B}^{(2)} (1,c,d,a,b,6 )+ A_{6:1B}^{(2)} (1,d,c,a,b,6 )
\end{eqnarray}
and act as consistecy contraints.    
For these we can solve for these combinations in terms of the other $SU(N_c)$ functions
\begin{equation}
A_{6:1B}^{sym_i} = \sum_{r,=1,3,4} \sum_i  c_{r,i}  A^{(2)}_{6:r} (\sigma_i)
\label{eq:con6B}\end{equation}
{\it but} these don't allow a solution for an individual $A^{(2)}_{6:1B}$.
The expressions of the RHS are subject to manipulation using eqs~(\ref{eq:DecCona}) and (\ref{eq:DecConb}).
A specific version are given in appendix~\ref{app:edison}.

After implementing these relations there remain only six relations amongst the rational terms of the all-plus amplitude.    This amplitude,   of course,   consists of more than rational terms.   If we also demand that the relations apply to the polylogarithmic parts also these six relations are reduced to a single relation which can be expressed
\begin{eqnarray}
A_{6:3} ^{sym}  ( a,b ; r,  s,  t, u )= 
& &A_{6:4} ^{sym}  (a,  b, r ; s,t,u ) +
A_{6:4} ^{sym}  (a,  b,  s ; r ,  t, u ) 
\notag 
\\
&+& A_{6:4} ^{sym}  (a,  b,  t ; r,  s,u ) +
A_{6:4} ^{sym}  (a,b,u ; r, s,t )  \; . 
\label{eq:extraid}
\end{eqnarray}
Although this looks like multiple relations by choosing different values for $(a,b,c,d,e,f)$ in fact it only adds one:  the difference between two choices of this relation lies within the space spanned by the decoupling identities.  

We can use (\ref{eq:extraid}) to obtain a new  
expression for $A_{6:2,2}^{(2)} $
which we can express as, 
\begin{eqnarray}
A_{6:2,2}^{(2)}  ( &a,b  &  ;  e, f ; r,s )=
\frac{1}{2} \sum_{ \sigma\in COP(\{e,f\}, \{r,s\})}A_{6:3} ^{(2)}  ( a,b  ; \sigma)
\notag \\
&+& \frac{1}{2} \sum_{ \sigma\in COP(\{a,b\}, \{r,s\})}A_{6:3} ^{(2)}  ( e,f ; \sigma ) 
+ \frac{1}{2} \sum_{ \sigma\in COP(\{a,b\}, \{e,f\})}A_{6:3} ^{(2)}  ( r,s ; \sigma)
\; . 
\label{eq:new622}
\end{eqnarray}
This form of $A_{6:2,2}^{(2)}$ is different from that of eqn.~(\ref{eq:A622first })  and is {\it not} a consequence of decoupling.  It relies upon the relation (\ref{eq:extraid}) -or is equivalent to it. 

Equations (\ref{eq:new622})  and (\ref{eq:extraid})  are only proven to apply to the all-plus amplitude but can be conjectured to apply more generally : this will be explored in the next section.

In summary,    we have confirmed the expected results for identities between the amplitudes and  have identified that there is a maximum of a single potential  relation beyond these.     Although only a single relation this hints at more general relations at higher points.  This is reminiscent of the Kleiss-Kuif relations at tree level.   We explore a possible extension of this to higher points in the next section.

\section{Triple Trace Term}
\def\ul#1{#1}
Up to eight points, the triple trace terms of the two loop amplitude are determined from the decoupling identities as discussed in eqn.~(\ref{eq:An2r}).    However,  there is a further speculative  relationship for the triple trace which we propose for all $n$,
\begin{eqnarray}
A_{n:r,s}^{(2)}( \ul{\alpha} ;  \ul{\beta} ;  \ul{\gamma})
&=& 
\frac{(-1)^{|\gamma|}}{2}
\sum_{\sigma\in COP(\alpha,\gamma^T)} A_{n:s+1}^{(2)}( \beta ; \sigma)
+\frac{(-1)^{|\beta|}}{2}
\sum_{\sigma\in COP(\gamma,\beta^T)} A_{n:r+1}^{(2)}( \alpha ; \sigma)
\notag\\
&+&\frac{(-1)^{|\alpha|}}{2}
\sum_{\sigma\in COP(\beta,\alpha^T)} A_{n:n-r-s+1}^{(2)}( \gamma ; \sigma)
\label{eq:tripletrace}
\end{eqnarray}
provided $ |\alpha|, |\beta|,|\gamma|  > 1$, 
which expresses the triple trace in terms of double trace terms.  

This relationship lies outside the relationships of the decoupling identities.    It is speculative but is satisfied for
\begin{equation}
\ul{\alpha} =\{ \alpha_1,\alpha_2\} ,\;  \ul{\beta} =\{ \beta_1,\beta_2\} ,\;  
\end{equation}
for the all-plus six point amplitude, as seen in the previous section., where $\ul{\gamma}=\{ \gamma_1,\gamma_2\}$.

The seven-point all plus amplitude has also been calculated~\citep{Dalgleish}.
We have checked that the rational parts of this specific helicity amplitude where 
$\ul{\gamma}=\{ \gamma_1,\gamma_2,\gamma_3\}$ also satisfies this relation.   For $n=7$, after solving for the $U(N_c)$ specific partial amplitudes and $A_{7:2:2}^{(2)}$ the decoupling identities impose consistency conditions among the $SU(N_c)$ partial functions (excluding $A_{n:7B}^{(2)}$) with rank 105.   Relation~(\ref{eq:tripletrace}) lies outside these identities. Unlike $n=6$ it is not a single additional relation but adds 35 to the rank of identities.   

The expression is robust against collinear limits.   For sets of length two, there is no distinction between the set and its transpose due to cyclic symmetry so our six and seven point amplitudes do not determine this.   However,  we have chosen the ordering so that the amplitude satisfies reflection symmetry for all $n$.   Similarly,  the overall factors may have a more general form when the sets have arbitrary length.

This relation has corrections if any of the traces has  length one.
If two have length one,   $\ul{\alpha}=\{\alpha_1\},  \ul{\beta}=\{\beta_1\}$ the following variant holds
\begin{eqnarray}
A_{n:1,1}^{(2)}( \ul{\alpha} ; \ul{\beta} ;\ul{\gamma})
&=& 
\frac{(-1)^{|\alpha|}}{2}
\sum_{\sigma\in COP(\gamma,\alpha^T)} A_{n:2}^{(2)}( \beta ;  \sigma)
+\frac{(-1)^{|\beta|}}{2}
\sum_{\sigma\in COP(\gamma,\beta^T)} A_{n:2}^{(2)}( \alpha ;  \sigma)
\notag\\
&+&{(-1)^{|\alpha|}}
\sum_{\sigma\in COP(\beta,\alpha^T)} A_{n:3}^{(2)}(  \sigma ; \gamma)
\; , 
\label{eq:relsimp1}
\end{eqnarray}
whilst if only one has length one, 
$\ul{\alpha}=\{\alpha_1\}$, $
\ul{\beta}=\{\beta_1,\cdots,\beta_{s-1}\}$
$\ul{\gamma}=\{\gamma_1,\cdots,\gamma_{t-1}\}$
\begin{eqnarray}
A_{n:1,s-1}^{(2)}( \ul{\alpha} ;  \ul{\beta} ; \ul{\gamma})= 
-\sum_{\sigma\in COP(\gamma,\alpha^T)} A_{n:s}^{(2)}( \beta ; \sigma)
&&
-\sum_{\sigma\in COP(\beta,\alpha^T)} A_{n:s+1}^{(2)}( \sigma ; \gamma )
\; .
\label{eq:rela11}
\end{eqnarray}
(If $s=t$,   $\sigma$ and $\gamma$ as interchanged in the final term.)
Unlike eqn.~(\ref{eq:tripletrace}),   eqs.~(\ref{eq:relsimp1}) and (\ref{eq:rela11})  can be derived from decoupling identities.

\section{$A_{n:1B}^{(2)}$}

There exists a conjectured form of $A_{n:1B}^{(2)} $ for the all-plus amplitude.~\citep{Dunbar:2020wdh}
\begin{equation}
A_{n:1B}^{(2)} (1^+,2^+,\cdots n^+)
\end{equation}
with closed expressions for the rational part of this amplitude.   These are  included in appendix~\ref{app:alln}.   
We use  this amplitude to check for relations. 
This partial amplitude has been verified for six and seven points amplitudes and has the correct symmetries, factorisations and collinear limits.    
For eight and nine point the expression has been verified numerically~\citep{Kosower:2022bfv}.

We have verified, up to $n=12$ that these specific helicity amplitudes satisfy 
Kleiss-Kuif type relations,  
\begin{equation}
A_{n:1B}^{(2)} (1,\{\alpha\}, n, \{\beta\}) =  
(-1)^{|\beta|} \sum_{\sigma \in OP(\alpha ,\beta^T)}  
A_{n:1B}^{(2)}( 1 , \{\sigma\} , n)
\; .
\label{eq:KKamp1B}
\end{equation}

Secondly,  we have checked there are no further relations solely amongst these
partial amplitudes.  Specifically there are no constants $c_{\sigma}$ satisfying
\begin{equation}
\sum_{\sigma}  c_{\sigma} A_{n:1B}^{(2)}  ( 1, \sigma(2,\cdots n-1), n ) =0 
\; .
\end{equation}
So these are no relations for the all plus this rules out any possible helicity independent relations.

It is very tempting to conjecture that equation~(\ref{eq:KKamp1B}) holds for all-$n$ and all helicities but we have limited data at this point.  The relation in eqn.~(\ref{eq:KKamp1B}) could even suggest a relationship between the four and three trace terms of three loop Yang-mills - however without any possible checks at this point.

\section{Conclusions} 

This article has explored possible relationships among the partial amplitudes of two-loop gluon scattering. This has essentially been an ``experimental'' style search using the known analytic forms of partial amplitudes.   This approach is exhaustive :  no potential relations will be missed.  This search is limited by the very small number of amplitudes which are known in closed analytic form.     Nonetheless we can draw some conclusions and make some conjectures.   In particular, on the negative side, we can confirm that the relations such  as eqn.~(\ref{eq:Edi}) does not extend beyond the five point case.   However on the positive side we have relationships of eqn.~(\ref{eq:tripletrace}) and ~eqn~(\ref{eq:KKamp1B})
which we present as plausible conjectures.

We thank Adam Dalgleish, Warren Perkins and Joe Strong for helpful comments and access to the unpublished seven point rational terms.  
This work was supported by the UKRI Science and Technology Facilities Council (STFC) Consolidated Grant No. ST/T000813/1.

\appendix

\section{Permutation Sums}
\label{app:permutations}

In the expressions of this article we have summations which are over sets formed by the merger of two sets.   In particular  we have the sum over ordered products $OP(\alpha,\beta)$ and the summation
over the cyclic ordered permutations $COP(\alpha,\beta)$ where,  due to cyclic symmetry,
cyclically equivalent permutations  are factored out 
In this appendix we illustrate these by some examples.
 For sets $\alpha_r$ and $\beta_s$ of length $r$ and $s$ respectively these sets  contain the following number of elements
\begin{eqnarray}
|OP( \alpha_r,  \beta_s)| &=& \frac{(r+s)!}{(r)!(s)!}  
\\
|COP( \alpha_r,  \beta_s)| &=& \frac{(r+s-1)!}{(r-1)!(s-1)!}  
\; .
\end{eqnarray}

As examples,  for $\alpha=\{1\}$, $\beta=\{2, 3,4,5\}$, 
\begin{eqnarray}
OP( \alpha,\beta)  &=&\Bigl\{\{ 1,2,3,4,5 \},   \{ 2,1,3,4, 5 \} , \{ 2,3,1,4,5 \}, 
\{ 2,3,4,1,5 \},    \{ 2,3,4,5,1 \} \Bigr\}
\notag\\
COP( \alpha,\beta)  &=&\Bigl\{\{ 1,2,3,4,5 \},   \{ 2,1,3,4, 5 \} , \{ 2,3,1,4,5 \}, 
\{ 2,3,4,1,5 \}  \Bigr\}
\end{eqnarray}
where $\{ 2,3,4,5,1 \},$ is omitted from $COP( \alpha,\beta)$ since it is cyclically equivalent to $\{ 1,2,3,4,5 \},$.

For $\alpha=\{1,2\}$, $\beta=\{3,4,5\}$, 
\begin{eqnarray}
OP( \alpha,\beta)  =\Bigl\{ 
\{ 1,2,3,4,5 \},  
\{ 1,3,2,4,5 \},  
\{ 1,3,4,2, 5 \},  
\{ 1,3,4,5,2 \},  
\notag\\
\{ 3,1,2,4,5 \},  
\{ 3,1,4,2,5 \},  
\{ 3,1,4,5,2 \},  
\{ 3,4, 1,2,5 \},  
\notag\\
\{ 3,4,1,5,2 \},  
\{ 3,4,5,1,2 \}  \Bigr\}
\end{eqnarray}
and
\begin{eqnarray}
COP( \alpha,\beta)  =\Bigl\{ \{ 1,2,3,4,5 \},  \{ 1,2,4,5,3 \},  \{ 1,2,5,3,4 \}, 
\{ 1,3,2,4,5 \}, 
\notag\\
\{ 1,4,2,5,3 \},   
\{ 1,5,2,3,4 \},   
\{1,3,4,2,5 \},\
\{1,4,5,2,3 \},
\notag\\
\{1,5,3,2,4 \},
\{1,3,4,5,2\},
\{1,4,5,3,2\},\{1,5,3,4,2\}   \Bigr\}  \; .
\end{eqnarray}

We can use this notation to re-express some of the formulae in this article. For example eqn.~(\ref{eq:decotree})
\begin{equation}
\sum_{\sigma \in COP(\{1\},\{2,3,\cdots n\}}
A^{(0)}_{n}(\sigma) =0 
\end{equation}
and eqn.~(\ref{eq:decoloop}) is
\begin{equation}
A^{(1)}_{n:2}(1 ;  2,3,\cdots n )+  \sum_{\sigma \in COP(\{1\},\{2,3,\cdots n\}}
A^{(1)}_{n:1}(\sigma) =0 
\; .
\end{equation}

\section{Explicit Form  for Constraints involving $A_{6:1B}^{(2)}$.}
\label{app:edison}
We provide here an explicit realisation of the constraints upon the six-point amplitudes involving $A_{6:1B}^{(2)}$ as in eqn.~(\ref{eq:con6B}),
\begin{eqnarray}
A_{6:1B}^{sym_1}
&\equiv&  
\sum_{sym(b,c,d)} A_{6:1B}^{(2)} (1,2,b,c,d,6 )
= \sum_{P(b,c,d)} \biggl( 
A_{6:4}^{sym}  ( 1, b, c ;  2,d,6 ) 
\notag \\
&+&  A_{6:3}^{(2)} ( 1, b ;   2, c , 6 , d )/2
-A_{6:3}^{(2)}  ( b, c ; 1,2,6 , d )/2
-A_{6:3}^{(2)}  ( 1, b ; 1, 6, 2 , d )/2  
\notag \\
&-&A_{6:1}^{(2)} ( 1,b,2,c,d,6) 
-A_{6:1}^{(2)}(1,2,b,c,6,d) 
+3 A_{6:1}^{(2)}(1,2,6,b,c,d)
\notag \\
&+&3 A_{6:1}^{(2)}(1,b 2,6,c,d)
+3 A_{6:1}^{(2)}(1,b,c,,d,2,6 )
+3 A_{6:1}^{(2)}(1,b,c,2,6, d)  \biggr)
\;  , 
\end{eqnarray}

\begin{eqnarray}
A_{6:1B}^{sym_3}
&\equiv &  
A_{6:1B}^{(2)} (1,a,b,c,d,6 )+
 A_{6:1B}^{(2)} (1,a,c,b,d,6 )\
% \notag\\ & &
   +
 A_{6:1B}^{(2)} (1,d,b,c,a,6 )+
 A_{6:1B}^{(2)} (1,d,c,b,a,6 ) 
 \notag\\
 &= &  
%TermF=
-A_{6:3}^{(2)}({1,6 ;  a, b, d, c})-A_{6:3}^{(2)}({ b, c ; 1, a,6, d})
%TermC=
-  A_{6:4}^{sym}({1, a, d ; b,c,6})- A_{6:4}^{sym}({1, b, c ; a,d,6})
%TermB=
\notag\\
&-& A_{6:4}^{sym}({1, a,6 ; b,c,d })- A_{6:4}^{sym}({1, b,6 ; a,c,d })
- A_{6:4}^{sym}({1, c,6 ; a,b,d})- A_{6:4}^{sym}({1, d,6; a,b,c})
\notag\\
 &+&  
  \sum_{b \leftrightarrow c} \Biggl(
A_{6:3}^{(2)}({1, b ;  a, c,6, d})+A_{6:3}^{(2)}({1, b ;  a, d, c,6})
+A_{6:3}^{(2)}({ b,6 ; 1, a, d, c})+A_{6:3}^{(2)}({ b,6 ; 1, c, a, d})   \Biggr)
 \notag\\
 &+ &   \sum_{a\leftrightarrow d , b \leftrightarrow c} \Biggl(
 A_{6:4}^{sym}(1, a, b ;  c, d,6 )
+A_{6:1}^{(2)}({1, a, b, c, d,6})
+A_{6:1}^{(2)}({1, a, b,6, d, c})
+A_{6:1}^{(2)}({1, a,6, b, d, c})
\notag\\
& &-3A_{6:1}^{(2)}({1, a, d, b, c,6})
-3A_{6:1}^{(2)}({1, a, d, b,6, c})
-3A_{6:1}^{(2)}({1, a, d,6, b, c})
\notag\\
& & -3A_{6:1}^{(2)}({1, b, a, d, c,6})
-3A_{6:1}^{(2)}({1, b, c, a, d,6})
-3A_{6:1}^{(2)}({1, b, a, d,6, c}) 
\Biggr) , 
\label{eq:ident6B3}
\end{eqnarray} 
 and
\begin{eqnarray}
A_{6:1B}^{sym_4}
&\equiv&  
 A_{6:1B}^{(2)} (1,2,3,4,5,6 )+ A_{6:1B}^{(2)} (1,2,3,5,4,6 )
+ A_{6:1B}^{(2)} (1,4,2,3,5,6 )
 \notag \\ 
 & & 
 + A_{6:1B}^{(2)} (1,5,2,3,4,6 )
 + A_{6:1B}^{(2)} (1,4,5,2,3,6 )+ A_{6:1B}^{(2)} (1,5,4,2,3,6 )
 \notag\\
  & =&  
 %
% termD=
 -A^{(2)}_{6:3} ({1, 2 ; 3, 4, 6, 5}) - A^{(2)}_{6:3} ({1, 6; 2, 4, 3, 5}) 
       -A^{(2)}_{6:3} ({2, 3 ;1, 4, 6, 5})
 \notag\\
& &       - A^{(2)}_{6:3} ({3, 6 ; 1, 4, 2, 5})
        -A^{(2)}_{6:3} ({4, 5 ; 1, 2, 6, 3}) - A^{(2)}_{6:3} ({4, 5 ;1, 3, 2, 6})
 \notag\\
%termA ( a_, b_]:=
&+&\sum_{a\leftrightarrow b} \Biggl( 2 A^{(2)}_{6:4} ({1,3, a ; 2, b,6})+ 2 A^{(2)}_{6:4} ({1,3, a ; 2,6, b})  
 +A^{(2)}_{6:4} ({1, 2,  a ; 3, 6,  b}) 
+A^{(2)}_{6:4} ({1,  a, 6 ; 2, 3,  b}) %   
  \notag\\    
%%t ermB ( a_, b_]:=
& & 
-A^{(2)}_{6:4} ({1,2,3 ;  a, b,6}) 
        -A^{(2)}_{6:4} ({1,2,6 ; 3, a, b})
         -A^{(2)}_{6:4} ({1,3,6 ; 2, a, b})
         -A^{(2)}_{6:4} ({1, a, b ; 2,3,6})     
 \notag\\
%termE ( a_, b_]:=(
& & +A^{(2)}_{6:1} ({1,  a, 2,  b, 3, 6}) 
+A^{(2)}_{6:1} ({1,  a, 2, 3, 6,  b})
+A^{(2)}_{6:1} ({1, 2, 3,  a, 6,  b})
+A^{(2)}_{6:1} ({1, 2,  a, 3,  b, 6})
 \notag\\
%termF ( a_, b_]:=(
& & -A^{(2)}_{6:1} ({1,  a, 2, 3,  b, 6}) 
-A^{(2)}_{6:1} ({1,  a,  b, 2, 3, 6}) 
-A^{(2)}_{6:1} ({1, 2, 3,  a,  b, 6}) 
 \notag\\
& & -A^{(2)}_{6:1} ({1, 2, 3, 6,  a,  b}) +
-A^{(2)}_{6:1} ({1, 2,  a, 3, 6,  b}) +
-A^{(2)}_{6:1} ({1, 2,  a,  b, 3, 6})
 \notag\\
%termG ( a_, b_]:=(
& &+6 A^{(2)}_{6:1} ({1, 3,  a, 2, 6,  b})
+6 A^{(2)}_{6:1} ({1, 3,  a, 6, 2,  b}) 
 \notag\\
%termH ( a_, b_]:=
& & +2 A^{(2)}_{6:1} ({1, 2,  a,  b, 6, 3})
+2 A^{(2)}_{6:1} ({1, 2,  a, 6, 3,  b})
+2 A^{(2)}_{6:1} ({1, 2,  a, 6,  b, 3})
+2 A^{(2)}_{6:1} ({1, 2, 6, 3,  a,  b})
 \notag\\
& & +2 A^{(2)}_{6:1} ({1, 2, 6,  a, 3,  b})
+2 A^{(2)}_{6:1} ({1, 3, 2,  a,  b, 6}) 
+2 A^{(2)}_{6:1} ({1, 3, 2,  a, 6,  b})
+2 A^{(2)}_{6:1} ({1, 3, 2,  b,  a, 6})
 \notag\\
& &+2 A^{(2)}_{6:1} ({1, 3,  a, 2,  b, 6})
+2 A^{(2)}_{6:1} ({1, 3, 6,  a, 2,  b})
+2 A^{(2)}_{6:1} ({1,  a, 2, 6, 3,  b}) 
+2 A^{(2)}_{6:1} ({1,  a, 3, 2,  b, 6})
 \notag\\
& &+2 A^{(2)}_{6:1} ({1,  a, 3, 2, 6,  b})
+2 A^{(2)}_{6:1} ({1,  a, 3,  b, 2, 6})
+2 A^{(2)}_{6:1} ({1,  a,  b, 3, 2, 6})
)
 \notag\\
%termI ( a_, b_]:=(
& &-2 A^{(2)}_{6:1} ({1, 2, 6,  a,  b, 3})
-2 A^{(2)}_{6:1} ({1, 3, 2,  a,  b, 6})
 \notag\\
& & -2 A^{(2)}_{6:1} ({1, 3, 2, 6,  a,  b})
-2 A^{(2)}_{6:1} ({1, 3,  a,  b, 2, 6})
-2 A^{(2)}_{6:1} ({1, 3, 6, 2,  a,  b}) 
\Biggr)  \; . 
\label{eq:ident6B4}
\end{eqnarray}

There are 14  independent identities.   No single identity of the above completes this but for example the identities (\ref{eq:ident6B3}) and (\ref{eq:ident6B4})  combined do.

\section{Rational Terms of Two Loop All plus six-point amplitude amplitude}
\label{app:sixpt}
\def\Fcc{P^{(2)}}
The IR singular structure of a color partial amplitude is determined by general theorems~\citep{Catani:1998bh}. Consequently we can split the amplitude into
singular terms $U^{(2)}_{n:\lambda}$ and finite terms and further split the finite terms into the polylogarithmic parts $\Fcc_{n:\lambda}$ and the rational parts $R_{n:\lambda}^{(2)}$
\begin{eqnarray}
\label{definitionremainder}
A^{(2)}_{n:\lambda} =& U^{(2)}_{n:\lambda}
+ \Fcc_{n:\lambda}+R_{n:\lambda}^{(2)}\; .  +   {\mathcal O}(\epsilon)\, .
\label{eq:constrituents}
\end{eqnarray}
As the all-plus
tree amplitude vanishes, $U^{(2)}_{n:\lambda}$ simplifies considerably 
and is at worst $1/\epsilon^2$~\citep{Kunszt:1994np,Dunbar:2019fcq}. 
The partial amplitudes of the six-point all-plus amplitude have all been computed.     The leading in color term was calculated first in \citep{Dunbar:2016gjb} and 
subsequently confirmed by Badger et.al.~\citep{Badger:2016ozq}. 
It was later presented in an alternative 
form \citep{Dunbar:2017nfy}. The  
remaining partial amplitudes were computed in~\citep{Dalgleish:2020mof}.  We use the rational part of these amplitudes to test relationships.  These are presented here for completeness.  

$\mathbf{R_{6:1}^{(2)}}$
\begin{align}
% A_{6:1}(a,b,c,d,e,f)|_{\mathbb{Q}}=
R_{6:1}^{(2)}(a,b,c,d,e,f)&= \frac{i}{9}\sum_{\mathcal{P}_{6:1}}
  \frac{G_{6:1}^1+G_{6:1}^2+G_{6:1}^3+G_{6:1}^4+G_{6:1}^5}{\spa{a}.b\spa{b}.c\spa{c}.d\spa{d}.e\spa{e}.f\spa{f}.a}
%\end{align}
\intertext{where}
%\begin{align}
    G_{6:1}^1(a,b,c,d,e,f)&=\frac{s_{cd}s_{df}\la f|a\,P_{abc}|e\ra}
    {\spa{f}.e\,t_{abc}}
    +\frac{s_{ac}s_{cd}\la a|f\,P_{def}|b\ra}
    {\spa{a}.b\,t_{def}},
    \notag\\
    G_{6:1}^2(a,b,c,d,e,f)&=\frac{\spb{a}.b\spb{e}.f}{\spa{a}.b\spa{e}.f}
    \spa{a}.e^2\spa{b}.f^2
    +\frac12\frac{\spb{f}.a\spb{c}.d}{\spa{f}.a\spa{c}.d}
    \spa{a}.c^2\spa{d}.f^2,
    \notag\\
    G_{6:1}^3(a,b,c,d,e,f)&=\frac{s_{df}\spa{f}.a\spa{c}.d\spb{a}.c\spb{d}.f}{t_{abc}},
    \notag\\
    G_{6:1}^4(a,b,c,d,e,f)&=\frac{\la a|be|f\ra t_{abc}}{\spa{a}.f}
     \notag\\ %   \intertext{and}
    G_{6:1}^5(a,b,c,d,e,f)&=s_{fa} s_{bc} 
    + s_{ac} s_{be}
    + \frac52 s_{af} s_{cd}  
    -8 [a|bcf|a\ra 
    -8 [a|cde|a\ra 
    -\frac12 [a|cdf|a\ra 
    - \frac{11}2 [b|cef|b\ra
\label{eq:leadingrat}
\end{align}

$\mathbf{R_{6:3}^{(2)}}$
\begin{align}
% A_{6:3}^{(2)}(a,b;c,d,e,f)|_{\mathbb{Q}}&= 
% \notag\\
R_{6:3}^{(2)}(a,b;c,d,e,f)&=
\sum_{\mathcal{P}_{6:3}}
\Bigg[\frac{i}3\Big(H^1_{6:3}(a,b,c,d,e,f)-H^1_{6:3}(a,b,c,d,f,e)\Big)
\notag\\
&+\frac{i}3
\frac{\Big(G_{6:3}^2(a,b,c,d,e,f)+G_{6:3}^3(a,b,c,d,e,f)+G_{6:3}^4(a,b,c,d,e,f)\Big)}
{\spa{a}.b\spa{b}.c\spa{c}.a\spa{d}.e\spa{e}.f\spa{f}.d}
\notag\\
&+\frac{i}{12}
\frac{G_{6:3}^5(a,b,c,d,e,f)}
{\spa{a}.b\spa{b}.c\spa{c}.d\spa{d}.e\spa{e}.f\spa{f}.a}
\Bigg]
\label{eq:r63full}
\end{align}
where
\begin{align}
H_{6:3}^1(a,b,c,d,e,f)&=
\frac{G_{6:3}^1(a,b,c,d,e,f)}
{\spa{a}.b\spa{b}.c\spa{c}.d^2\spa{d}.e\spa{e}.f\spa{f}.a}
+ \frac{\spb{c}.d}{\spa{c}.d^2}
\frac{\spa{c}.f \spa{d}.b [b|f|d\ra}
{\spa{a}.b \spa{a}.f \spa{b}.f \spa{d}.e \spa{e}.f}
\notag\\
G_{6:3}^1(a,b,c,d,e,f)&= 
s_{ce} \la c|bf|d\ra - s_{cf} \la c|be|d\ra 
\notag\\
G_{6:3}^2(a,b,c,d,e,f)&=\frac{[d|P_{def}b|a]\la d|fP_{def}|a\ra 
+s_{de}[f|cbd|f\ra 
+[b|df|e]\la b|c P_{abc}|e\ra} 
 {t_{def}}
\notag\\
G_{6:3}^3(a,b,c,d,e,f)&=-\frac{s_{df} \la d|fb|c\ra [c|P_{abc}|e\ra}
{\spa{d}.e  t_{def}}
-
\frac{s_{de} \la f|db|c\ra [c|d|e\ra}
{\spa{e}.f t_{def}}
\notag\\
G_{6:3}^4(a,b,c,d,e,f)&=
-s_{bd}s_{de} - [a|bde|a\ra + [b|cde|b\ra 
-[a|bdf|a\ra 
\notag\\
&+[b|cdf|b\ra 
+[b|cef|b\ra 
-[b|def|b\ra
\notag\\
G_{6:3}^5(a,b,c,d,e,f)&=
-4s_{ac}^2 + 2s_{ab}s_{ad} - 2s_{ac}s_{ad}
+2s_{ab}s_{ae} - 2s_{ac}s_{ae} + 2s_{bd}^2 - 2s_{be}^2
+2s_{bf}^2
\notag\\
&- 8s_{ac}s_{cd} + 4s_{bc}s_{cd} + 
 12 s_{bd}s_{cd} + 6 s_{cd}^2 - 8 s_{ac} s_{ce} + 
 12 s_{bc} s_{ce} + 16 s_{bd} s_{ce}
 \notag\\
 &+ 4 s_{be} s_{ce} + 
 8 s_{cd} s_{ce} + 2 s_{c e}^2 + 2 s_{c f}^2 - 8 s_{a c} s_{d e} - 
 4 s_{a d} s_{d e} - 4 s_{b c} s_{d e} + 4 s_{c d} s_{d e}
 \notag\\
 & +4 s_{c e} s_{d e} - 8 [a|bce|a\ra 
 - 39 [a|bcf|a\ra - 
 18 [a|bdf|a\ra + 2 [a|bef|a\ra 
 \notag\\
 &- 10 [a|cdf|a\ra  - 2 [a|cef|a\ra - 4 [a|def|a\ra + 8 [b|cde|b\ra - 
 4 [b|cdf|b\ra 
 \notag\\
 &- 4 [b|cef|b\ra - 4 [b|def|b\ra - 
 4 [c|def|c\ra
 \label{eq:r63pieces}
\end{align}
$\mathbf{R_{6:4}^{(2)}   }$
\begin{align}
% A_{6:4}^{(2)}(a,b,c;d,e,f)|_{\mathbb{Q}}&= 
% \notag\\
 R_{6:4}^{(2)}(a,b,c,d,e,f)&=\frac{i}{36}\sum_{\mathcal{P}_{6:4}}
    \Bigg[\frac{\Big(G_{6:4}^{1}(a,b,c,d,e,f)+G_{6:4}^{2}(a,b,c,d,e,f)\Big)
 }
 {\spa{a}.b\spa{b}.c\spa{c}.a\spa{d}.e\spa{e}.f\spa{f}.d}
 \notag\\
 &\hspace{1.3cm}+12\frac{\Big(G^{3}_{6:4}(a,b,c,d,e,f)+G^{4}_{6:4}(a,b,c,d,e,f)\Big)}{\spa{a}.b\spa{c}.d\spa{d}.e\spa{e}.f\spa{f}.c}
 \Bigg],
\end{align}
 where
 \begin{align}
 G^{1}_{6:4}(a,b,c,d,e,f)&=\frac{4\,\la e|P_{abc}a|b\ra[e|dP_{abc}|b]}{t_{abc}},
 \notag\\
 G^{2}_{6:4}(a,b,c,d,e,f)&= s_{ad}^2 + 106\,s_{ab}s_{ad} + 102\,[a|bcd|a\ra - 4\,[a|bde|a\ra - 4\,[a|dbe|a\ra,
 \notag\\
 G^{3}_{6:4}(a,b,c,d,e,f)&=-\frac{\spb{a}.b}{\spa{a}.b}\Bigl(\la a|cd|b\ra+\la a|ef|b\ra\Bigr),
 \notag\\
 G^{4}_{6:4}(a,b,c,d,e,f)&=[a|cd|b]+[a|ef|b].
\end{align}
$\mathbf{R_{6:2,2}^{(2)}}$
\begin{align}
% A_{6:2,2}(a,b;c,d;e,f)|_{\mathbb{Q}} &= 
% \notag\\
R_{6:2,2}^{(2)}(a,b;c,d;e,f)&=\sum_{\mathcal{P}_{6:2,2}}
    i\,\frac{G_{6:2,2}^{1}(a,b,c,d,e,f)+G_{6:2,2}^{2}(a,b,c,d,e,f)
 }
 {\spa{a}.b\spa{b}.c\spa{c}.a\spa{d}.e\spa{e}.f\spa{f}.d},
\end{align}
where
\begin{align}
G_{6:2,2}^{1}(a,b,c,d,e,f)&= \frac{\la b|P_{abc}f|d\ra[b|cP_{abc}|d] } 
 {t_{abc}},
 \notag\\
 G_{6:2,2}^{2}(a,b,c,d,e,f)&=s_{ad}[e|P_{bc}|e\ra -s_{ac}[e|P_{fa}|e\ra -s_{af}s_{ae} -s_{ae}s_{cd}.
\end{align}

\section{All $n$ form of $R_{n:1B}^{(2)} (1^+,2^+,\cdots n^+)$}
\label{app:alln}

There exists a conjectured form of $A_{n:1B}^{(2)} $ for the all-plus amplitude.~\citep{Dunbar:2020wdh}.   We use  this amplitude to check the relation of eqn.(\ref{eq:KKamp1B}).  
The amplitude has been verified for six and seven points amplitudes and has the correct symmetries, factorisations and collinear limits.    For eight and nine point the expression has been verified numerically~\citep{Kosower:2022bfv}.
The amplitude is split into its constituents as in eqn.~(\ref{eq:constrituents}).    The rational term is split into two parts
\begin{eqnarray}
  R^{(2)}_{n:1B}(1^+,2^+,\cdots ,n^+) 
 & &
  =  R^{(2)}_{n:1B_1}(1^+,2^+,\cdots ,n^+) 
  +
  R^{(2)}_{n:1B_2}(1^+,2^+,\cdots ,n^+) 
\end{eqnarray}
where
\begin{equation}
 R^{(2)}_{n:1B_1}(1^+,2^+,\cdots ,n^+) =-2i \, \text{C}_{\text{PT}} (1,2,\cdots ,n-1,n) \times\sum_{1\leq i < j < k < l \leq n } \epsilon(i,j,k,l)
\end{equation}
\begin{eqnarray}
 & & R^{(2)}_{n:1B_2}(1^+,2^+,\cdots ,n^+) =4i \sum_{r=1}^{n-4}\sum_{s=r+4}^n
\notag \\
& & 
 \sum_{i=r+1}^{s-2}\sum_{j=i+1}^{s-1}
\epsilon(\{1,\cdots, r\},j,i,\{s,\cdots , n\} ) (-1)^{i-j+1}
\times \sum_{\alpha\in S_{r,s,i,j}}  \text{C}_{\text{PT}} (\{\alpha_{S_{r,s,i,j}} \} ) \,.
\end{eqnarray}

We have defined
\begin{equation}
\epsilon(\{a_1,a_2,\cdots, a_m \}, b , c , \{d_1,d_2,\cdots, d_p\})
\equiv\sum_{i=1}^{m}\sum_{j=1}^p \epsilon(a_i , b , c , d_j)\,,
\end{equation}
\begin{equation}
\PT (a_1,a_2,a_3,\cdots, a_n ) \equiv { 1\over \spa{a_1}.{a_2}\spa{a_2}.{a_3} \cdots \spa{a_n}.{a_1} }  
\equiv{1\over \CY( a_1,a_2,a_3,\cdots, a_n  )} \,.
\end{equation}
To define $S_{r,s,i,j}$ we divide the list of indices, 
\begin{align}
\{ 1,2,3,\cdots , n\} &=  \{ 1, \cdots ,r; r+1, \cdots, i-1; i ; i+1, \cdots , j-1; j; j+1,\cdots ,s-1; s,\cdots, n \}
\notag \\
&\equiv\{ 1, \cdots r, \} \oplus S_1 \oplus\{ i \} \oplus S_2  \oplus\{ j \} \oplus  S_3 \oplus \{ s, \cdots, n \}
\end{align}
with
\begin{equation}
S_1=\{r+1, \cdots ,i-1\}, \;\;\; S_2=\{  i+1, \cdots ,j-1\} , \;\;\ S_3=\{ j+1,\cdots ,s-1 \} \,.
\end{equation}
The sets $S_i$ may be null. Then 
\begin{equation}
S_{r,s,i,j} =  Mer( S_1, \bar S_2, S_3) 
\end{equation}
where $\bar S_2$ is the reverse of $S_2$ and $Mer( S_1, \bar S_2, S_3)$ is the set of all mergers of the three sets which respect the ordering
within the $S_i$ 
and
\begin{equation}
\alpha_{S_{r,s,i,j}} = \{ 1, \cdots, r \} \oplus \{ j \} \oplus \alpha \oplus \{ i \} \oplus \{ s, \cdots, n \}
\; . 
\end{equation}

\bibliography{TwoLoop}{}

\bibliographystyle{JHEP}

\end{document}

\end{thebibliography}
\end{document}